\documentclass[aps,prb,twocolumn,showpacs,amsmath,amssymb,superscriptaddress]{revtex4-1}

\usepackage{amsmath,amsfonts,bm,graphicx,verbatim,mathrsfs,units}
\usepackage[colorlinks=true,citecolor=blue]{hyperref}

\newcommand{\bra}[1]{\langle #1|}
\newcommand{\EE}{{\cal S}}
\newcommand{\eps}{\varepsilon}
\newcommand{\ew}[1]{\langle #1\rangle}

\newcommand{\hatc}{\hat{c}}
\newcommand{\hatH}{\hat{H}}
\newcommand{\hatM}{\hat{M}}
\newcommand{\hatn}{\hat{n}}
\newcommand{\hatP}{\hat{P}}
\newcommand{\hatrho}{\hat{\rho}}
\newcommand{\hatU}{\hat{U}}
\newcommand{\hc}{\textrm{h.c.}}
\newcommand{\ket}[1]{|#1\rangle}
\newcommand{\of}[1]{\left(#1\right)}
\newcommand{\tb}{\bar{t}}
\newcommand{\tr}{\textrm{Tr}}
\newcommand{\tq}{t_\mathrm{QPC}}

\begin{document}
\title{Entanglement entropy in dynamic quantum-coherent conductors}

\author{Konrad H.\ Thomas}
\affiliation{D\'epartement de Physique Th\'eorique, Universit\'e de Gen\`eve, 1211 Gen\`eve, Switzerland}
\author{Christian Flindt}
\affiliation{D\'epartement de Physique Th\'eorique, Universit\'e de Gen\`eve, 1211 Gen\`eve, Switzerland}
\affiliation{Department of Applied Physics, Aalto University, 00076 Aalto, Finland}
\date{\today}

\begin{abstract}
We investigate the entanglement and the R\'enyi entropies of two electronic leads connected by a quantum point contact. For non-interacting electrons, the entropies can be related to the cumulants of the full counting statistics of transferred charge which in principle are measurable. We consider the entanglement entropy generated by operating the quantum point contact as a quantum switch which is opened and closed in a periodic manner. Using a numerically exact approach we analyze the conditions under which a logarithmic growth of the entanglement entropy predicted by conformal field theory should be observable in an electronic conductor. In addition, we consider clean single-particle excitations on top of the Fermi sea (levitons) generated by applying designed pulses to the leads. We identify a Hong-Ou-Mandel-like suppression of the entanglement entropy by interfering two levitons on a quantum point contact tuned to half transmission.
\end{abstract}

\pacs{03.67.Mn, 72.70.+m, 73.23.-b}


\maketitle

\section{Introduction}
\label{sec:Introduction}

The concept of entanglement entropy is currently at the forefront of condensed matter physics.\cite{Amico2008,Horodecki2009,Eisert2010} Originally developed in the context of black hole physics,\cite{Bombelli1986} entanglement entropy was later adopted in the quantum information sciences to quantify the degree of entanglement between two parties.\cite{Bennett1996} In recent years, it has also been recognized as a useful quantity in condensed matter systems, for instance to investigate quantum critical systems,\cite{Vidal2003,Lambert2004,Refael2004,Calabrese2004,Metlitski2009,Kallin2009} quantum quenches,\cite{Calabrese2005,Bravyi2006,Eisert2006, Calabrese2007,Eisler2007,Cramer2008} topologically ordered states, \cite{Kitaev2006,Levin2006,Dong2008} and strongly correlated systems.\cite{Verstraete2004}
One important finding is that in gapless one-dimensional fermionic systems, the entanglement entropy depends logarithmically on the system size,\cite{Holzhey1994}
while in quenched systems the role of spatial extent is played by time.\cite{Calabrese2004,Eisler2007}

Despite the theoretical interest, the measurement of entanglement entropy remains challenging, since its definition does not refer to any measurable observables. This has prompted a search for schemes to measure the entanglement entropy in quantum many-body systems.\cite{Klich2006,Cardy2011,Abanin2012,Daley2012,Pichler2013,Klich2009,Song2011,Song2012, Petrescu2014} One proposal\cite{Klich2009,Song2011,Song2012,Petrescu2014} relates the entanglement entropy between two reservoirs of non-interacting electrons to the full counting statistics (FCS) of transferred charge.\cite{Levitov1993,Levitov1996,Nazarov2003} In this approach, first suggested by Klich and Levitov\cite{Klich2009} and later refined in Refs.~\onlinecite{Song2011,Song2012}, the entanglement and the R\'enyi entropies are expressed as series in the cumulants of the FCS. Since charge fluctuations in nano-scale electronics are now being detected experimentally,\cite{Reulet2003,Bomze2005,Fujisawa2006,Gustavsson2006,Timofeev2007,Sukhorukov2007,Gershon2008, Gabelli2009,Flindt2009,Gustavsson2009,Maisi2011,Ubbelohde2012,Maisi2014} these relations may provide a means to measure the entanglement entropy in quantum-coherent conductors.

\begin{figure}
	\includegraphics[width=0.95\columnwidth]{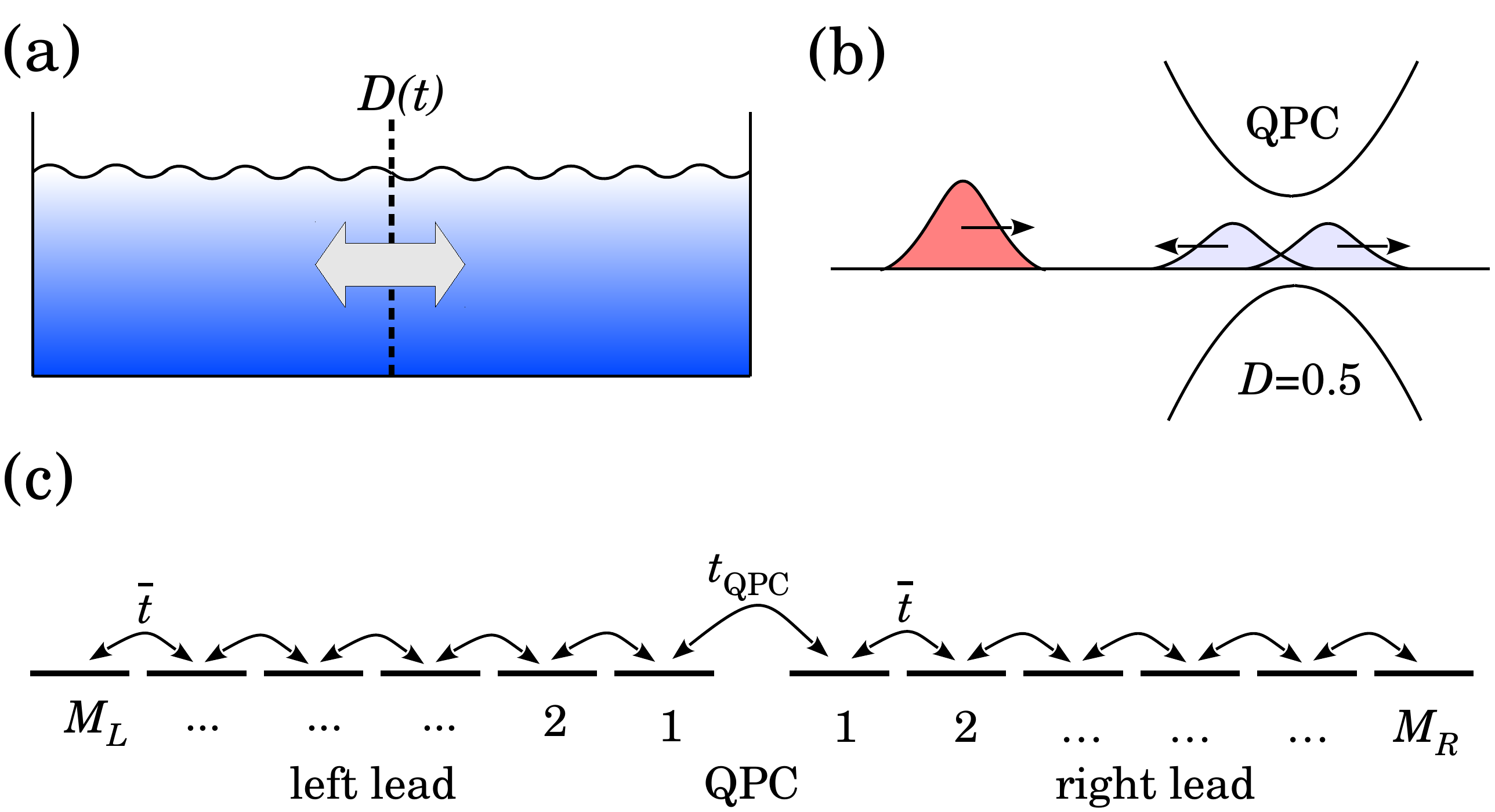}
	\caption{\label{Fig:Fig1} (Color online) Entanglement entropy in quantum-coherent conductors. (a) Two Fermi seas connected via a barrier with a time-dependent transmission $D(t)$. The reservoirs exchange particles, leading to many-body entanglement of the Fermi seas. (b) The barrier may consist of a quantum point contact (QPC) connecting two nano-scale electrodes. Here, a clean single-particle excitation (a leviton) is partitioned on the QPC tuned to half transmission. (c)  Tight-binding model of two leads connected by the tunneling amplitude $\tq$. Each lead consists of $M_{L/R}$ sites with tunneling amplitude $\tb$.}
\end{figure}

In this work we investigate the entanglement and the R\'enyi entropies in dynamic conductors making use of the connections to FCS.\cite{Klich2009,Song2011,Song2012,Petrescu2014}  We consider the schematic setup in Fig.~\ref{Fig:Fig1}a, showing two Fermi seas connected by a constriction whose transmission can be controlled in a time-dependent manner. To be specific, such a setup can be realized experimentally by connecting two electronic leads via a quantum point contact (QPC) as illustrated in Fig.~\ref{Fig:Fig1}b. We concentrate on time-dependent situations,\cite{Vanevic2007,Vanevic2008} where either the transmission of the QPC is modulated in time, or designed pulses are applied to the leads to generate clean single-particle excitations (levitons) on top of the Fermi sea, following the proposal by Levitov and co-workers\cite{Levitov1993,Levitov1996,Keeling2006} and the recent breakthrough-experiments reported in Refs.~\onlinecite{Dubois2013nature,Jullien2014}.

We are interested in a setup where the QPC is operated as a quantum switch that is opened and closed in a periodic manner. In this case, we show that a logarithmic growth of the entanglement entropy --- as predicted by conformal field theory\cite{Holzhey1994,Vidal2003,Calabrese2004} --- should be observable under suitable experimental conditions. Importantly, as we find, the logarithmic growth can be inferred from a measurement of only the first few cumulants of the FCS. We also consider the entanglement entropy produced by partitioning levitons on the QPC, as shown in Fig.~\ref{Fig:Fig1}b, and by interfering two levitons on the QPC tuned to half transmission. In this case, we identify a Hong-Ou-Mandel-like\cite{Hong1987,Bocquillon2013} suppression of the entanglement entropy as a function of the difference in arrival times at the QPC.

Problems concerning finite-time FCS for time-dependent systems are difficult to treat analytically. Instead, we employ a numerically exact scheme based on the tight-binding model in Fig.~\ref{Fig:Fig1}c. The FCS has previously been investigated for such systems without an external driving.\cite{Schonhammer2007,Inhester2009,Schonhammer2009,Levine2012} Here we extend the approach to time-dependent Hamiltonians (see also Refs.~\onlinecite{Zhang2009,Sherkunov2009,Jonckheere2012}). With this method, we may investigate the influence of the external modulations on the cumulants and the entanglement entropy as functions of time, and we can identify the number of cumulants needed in an experiment to reliably approximate the entanglement entropy. We focus here on mesoscopic conductors, but our tight-binding model may also describe cold atoms in optical lattices.\cite{Brantut2012,Krinner2014,Chien2012a,Chien2012b,Chien2014}

The rest of the paper is structured as follows. In Sec.~\ref{sec:Formalism} we introduce the entanglement and the R\'enyi entropies and reiterate how for non-interacting fermions they can be expressed as series in the cumulants of the FCS. We illustrate these ideas with a simple example involving only two fermions. In this connection, we also discuss the concept of accessible entanglement. In Sec.~\ref{sec:TB-model} we describe our tight-binding model and show how the entropies and the cumulants of the FCS can be evaluated. In Sec.~\ref{sec:Switch} we then consider the quantum switch. We investigate the increase of the entanglement entropy upon opening the QPC and how it is affected by finite temperatures, finite bias, and a non-perfect transmission. This problem can be addressed with our numerically exact scheme, making no further approximations. In Sec.~\ref{sec:Levitons} we consider clean single-particle excitations above the Fermi sea (levitons) generated by applying
designed pulses to the leads. In this case, we identify a Hong-Ou-Mandel-like suppression of the entanglement entropy by interfering two levitons on the quantum point contact. Finally, in Sec.~\ref{sec:Conclusion} we summarize our results. Technical details are provided in the appendices.

\section{Formalism}
\label{sec:Formalism}

\subsection{Entanglement entropy}
\label{sec:EE}

The entanglement entropy of a quantum many-body system is defined with respect to a partitioning of the system into a subsystem and its complement. Here we analyze the entanglement entropy between particles in two electronic leads connected by a QPC as illustrated in Fig.~\ref{Fig:Fig1}a. To define the entanglement entropy of the right lead ($R$) we introduce the reduced density matrix
\begin{equation}
\hatrho_R=\tr_L[\hatrho],
\end{equation}
obtained by tracing out the degrees of freedom in the left lead ($L$). The density matrix of the full system is denoted as $\hatrho$. An analogous definition holds for the left lead.

The entanglement entropy of the right lead is defined as the von Neumann entropy of $\hatrho_R$,\cite{Amico2008,Horodecki2009,Eisert2010}
\begin{equation}
\EE_R=-\tr_R[\hatrho_R\ln\hatrho_R],
\label{eq:EE_def}
\end{equation}
and analogously for the left lead. If the full system is in a pure state $\hatrho=|\Psi\rangle\!\langle\Psi|$, it holds that $\EE_R=\EE_L$.

A larger class of entanglement measures is provided by the R\'enyi entropies\cite{Renyi1960}
\begin{equation}
\EE_R^{(\nu)}=\frac1{1-\nu}\ln\of{\tr_R[\hatrho_R^\nu]}
\label{eq:RE_def}
\end{equation}
of order $\nu$. The entanglement entropy can be obtained from the limit $\nu\to1$. In the following we treat identical leads such that $\EE^{(\nu)}_L=\EE^{(\nu)}_R$. We thus skip the subscript $L$ or $R$ and evaluate the entropies in the right lead. For a system with $N$ particles in a pure state, the entanglement entropy takes on values between zero for a product state and $N\ln2$ for a maximally entangled state.\cite{Bennett1996}

\subsection{Full counting statistics}
\label{sec:FCS}

The definitions of the entanglement entropy and the R\'enyi entropies are general and can be applied to a variety of quantum systems. However, they do not refer to any directly measurable observables, making a measurement of the entropies a difficult task. In this work we consider non-interacting electrons.  The entanglement and the R\'enyi entropies can then be expressed in terms of the cumulants of the FCS. Specifically, the entanglement entropy is obtained as the limit\cite{Song2011,Song2012}
\begin{equation}
\EE=\lim_{K\to\infty}\EE_K
\label{eq:EEcumulantslimit}
\end{equation}
of the series
\begin{equation}
\EE_K=\sum_{m=1}^{K+1} a_m(K)C_m.
\label{eq:EEcumulants}
\end{equation}
The cut-off dependent coefficients are
\begin{equation}
a_m(K)=\left\{\begin{array}{cc}2\sum_{k=m-1}^K\frac{S_1(k,m-1)}{k!k} &m\textrm{ even},\\
0 & m\textrm{ odd}\end{array}\right.,
\end{equation}
where $S_1(n,m)$ are the unsigned Stirling numbers of the first kind. The cumulants $C_m$ are defined with respect to the probabilities $P_n$ that $n$ particles have been transferred between the leads during the time span $[t_0,t]$ with $t_0$ denoting the time at which the counting of particles begins. Taking the logarithm of the moment generating function
\begin{equation}
\chi(\lambda)=\sum_{n=-\infty}^\infty P_ne^{i\lambda n},
\label{eq:MGF_def}
\end{equation}
the cumulants follow by differentiation with respect to the counting field $\lambda$ at $\lambda=0$,
\begin{equation}
C_m=\frac{\partial^m}{\partial(i\lambda)^m}\ln\chi(\lambda)\big|_{\lambda=0}.
\label{eq:cumulants_def}
\end{equation}
The cumulants are time-dependent, since the probabilities $P_n$ depend on the length of the time interval $[t_0,t]$. The series (\ref{eq:EEcumulants})
provides an increasingly accurate lower bound to the exact entanglement entropy and it converges from below to the exact value as more cumulants are included.\cite{Song2011,Song2012}

The R\'enyi entropies can also be related to the cumulants of the FCS as the limits\cite{Song2012}
\begin{equation}
\EE^{(\nu)}=\lim_{K\to\infty}\EE_K^{(\nu)}
\end{equation}
of the series
\begin{equation}
\EE_K^{(\nu)}=\sum_{m=1}^K s_m^{(\nu)}C_m.
\label{eq:REcumulants}
\end{equation}
For the R\'enyi entropies of integer order, the coefficients $s_m^{(\nu)}$ are independent of the cutoff $K$ and read\cite{Calabrese2012}
\begin{equation}
s_m^{(\nu)}=\left\{\begin{array}{cc}
\frac{(-1)^{m/2}(2\pi)^m2\zeta[-m,(1+\nu)/2]}{(\nu-1)\nu^m m!} &m\textrm{ even},\\
0 & m\textrm{ odd}\end{array}\right. ,
\end{equation}
where $\zeta(s,a)=\sum_{n=0}^\infty (n+a)^{-s}$ is the generalized zeta function.

For non-interacting electrons, the FCS generally takes the form of a (discrete) generalized binomial distribution.\cite{Abanov2008,Abanov2009}
However, in some situations one may assume that the charge fluctuations are essentially gaussian\cite{Levitov1996} (a continuous distribution), so that only the first and second cumulants are non-zero. The entanglement entropy then becomes\cite{Klich2009}
\begin{equation}
{\cal S}\simeq\frac{\pi^2}{3}C_2,
\label{eq:EE_Gauss}
\end{equation}
having used the limiting value
\begin{equation}
\lim_{K\to\infty}a_2(K)=\frac{\pi^2}{3}
\end{equation}
for the prefactor. Similarly, the R\'enyi entropies read\cite{Song2011,Song2012}
\begin{equation}
{\cal S}^{(\nu)}\simeq s_2^{(\nu)}C_2
\label{eq:RE_Gauss}
\end{equation}
for gaussian fluctuations.

\subsection{Two particles}
\label{sec:two_particles}

\begin{figure}
	\includegraphics[width=0.95\columnwidth]{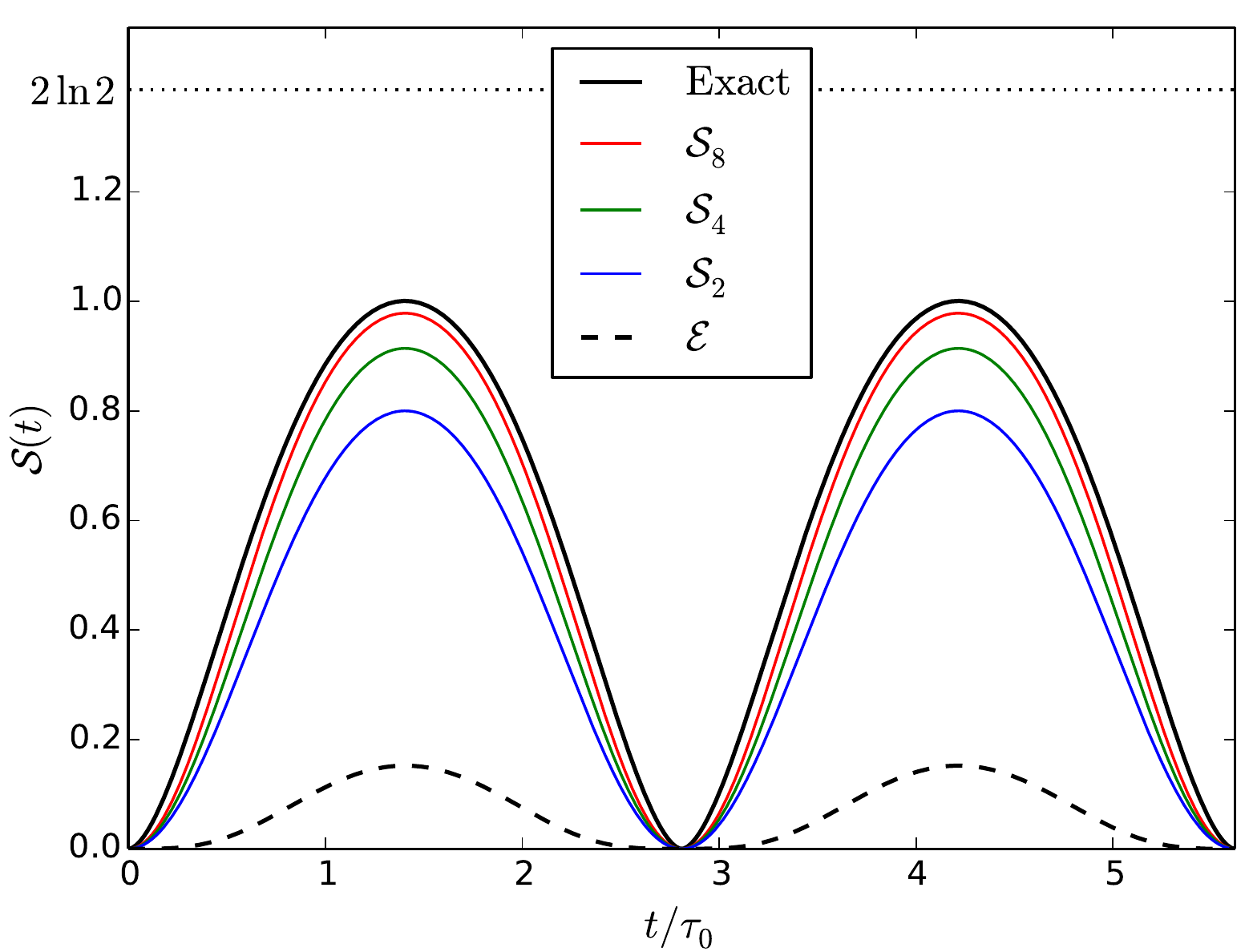}
	\caption{\label{Fig:EE_N=2}  (Color online) Entanglement entropy of two particles. The black line shows the exact result. The colored lines are lower bounds obtained by including an increasing number of cumulants in the series $\mathcal{S}_K$.  The maximal possible value of the entanglement entropy is shown with a dotted line. The dashed line shows the accessible entanglement entropy ${\cal E}$. The unit of time is $\tau_0=\hbar/\tb$.}
\end{figure}

Before discussing the details of our calculations, we consider a simple example where two non-interacting electrons get delocalized across the tight-binding chain in Fig.~\ref{Fig:Fig1}a with only $M_L=M_R=2$ sites in each lead. We initialize the system at $t_0=0$ in the groundstate of the uncoupled leads,
\begin{equation}
\ket{\psi(t=0)}=\frac12\left(\hatc_{1L}^\dag+\hatc_{2L}^\dag\right)
\left(\hatc_{1R}^\dag+\hatc_{2R}^\dag\right)\ket{0},
\label{eq:prod_state}
\end{equation}
where $\ket{0}$ is the vacuum state and $\hatc_{i\alpha}^\dag$ creates an electron at site $i$ ($=1,2$) in lead $\alpha$ ($=L,R$). This is a product state with zero entanglement entropy.

Next, we couple the two leads and let the system evolve with the Hamiltonian
\begin{equation}
\hat{\cal H}=-\tb\left(\hatc_{2L}^\dag \hatc_{1L}^{\;} +\hatc_{1L}^\dag \hatc_{1R}^{\;}+ \hatc_{1R}^\dag \hatc_{2R}^{\;}\right)+\hc,
\end{equation}
having taken $\tb=\tq$ in Fig.~\ref{Fig:Fig1}a. At the later time $t>0$, the state of the system takes the form
\begin{equation}
\begin{split}
\ket{\psi(t)}= \Big[  &a(t)\left(\hatc_{2L}^\dag\hatc_{1L}^\dag+\hatc_{1R}^\dag\hatc_{2R}^\dag\right)\\
&+b(t)\left(\hatc_{2L}^\dag\hatc_{1R}^\dag+\hatc_{1L}^\dag\hatc_{2R}^\dag\right)\\
&+c(t)\left(\hatc_{1L}^\dag\hatc_{1R}^\dag+\hatc_{2L}^\dag\hatc_{2R}^\dag\right)\Big]\ket{0}
\end{split}
\label{eq:state}
\end{equation}
where the explicit expressions for the coefficients $a(t)$, $b(t)$, and $c(t)$ are cumbersome and not shown here. The state is normalized such that
\begin{equation}
|\langle\psi(t)|\psi(t)\rangle|^2= 2\left(|a(t)|^2+|b(t)|^2+|c(t)|^2\right)=1
\end{equation}
at all times. The coefficient $a(t)$ in Eq.~(\ref{eq:state}) multiplies terms with either zero or two electrons in the right lead. The other terms correspond to entangled states with one electron in each lead.

To evaluate the entanglement entropy, we trace out the degrees of freedom of the left lead to find the reduced density matrix of the right one,
\begin{equation}
\label{eq:den_mat_2}
\hatrho_R=\left(\begin{array}{cccc} |a|^2 & 0 & 0 & 0 \\ 0 & |b|^2+|c|^2 & bc^*+b^*c & 0 \\ 0 & b^*c+bc^* & |b|^2+|c|^2 & 0 \\ 0 & 0 & 0 & |a|^2 \end{array}\right),
\end{equation}
in the basis $\{\ket{0}, \hatc_{1R}^\dag\ket{0}, \hatc_{2R}^\dag\ket{0}, \hatc_{1R}^\dag\hatc_{2R}^\dag\ket{0}\}$. By diagonalizing $\hatrho_R$, we obtain the entanglement entropy from the definition (\ref{eq:EE_def})
\begin{equation}
\EE=2g(|a|^2)+g(|b+c|^2)+g(|b-c|^2),
\end{equation}
having introduced the function $g(x)=-x\ln x$.

In Fig.~\ref{Fig:EE_N=2} we show the entanglement entropy as a function of time. The system oscillates back and forth between the product state in Eq.~(\ref{eq:prod_state}) and an entangled state with $\EE\simeq 1$. Together with the exact results for the entanglement entropy, we show results obtained from the series (\ref{eq:EEcumulants}) with an increasing number of cumulants included. (The details of these calculations are presented below.) The series converges to the exact results, showing that the entanglement entropy can be obtained from a measurement of the FCS.

It should be noted that conservation laws may restrict the amount of accessible entanglement. In our case this concerns particle conservation as entanglement between states with different particle numbers generally is not considered useful due to superselection rules.\cite{Wiseman2002,Klich2008}  To account for this, one may consider the individual (normalized) density matrices $\hatrho_R^{(N)}$ for each subspace with a fixed particle number ($N=0,1,2$ in our example). The accessible entanglement ${\cal E}$ is then the sum of the entropies of each density matrix, weighted with the probability $P_N$ of finding $N$ particles in the subsystem,\cite{Wiseman2002} i.~e.
\begin{equation}
{\cal E}=-\sum_N P_N\tr\left[\hatrho_R^{(N)}\ln\hatrho_R^{(N)}\right].
\end{equation}
For our system, we find
\begin{equation}
{\cal E}=g\of{|b+c|^2}+g\of{|b-c|^2}-g\of{2(|b|^2+|c|^2)},
\end{equation}
which is also shown in Fig.~\ref{Fig:EE_N=2}. The accessible entanglement is substantially smaller than the entanglement entropy. However, it has been shown that the difference between $\EE$ and $\cal E$ in many cases grows only logarithmically with the entanglement entropy and therefore is unimportant when many particles are involved.\cite{Klich2008}

\section{Tight-binding model}
\label{sec:TB-model}

\begin{figure*}
\includegraphics[width=0.95\textwidth]{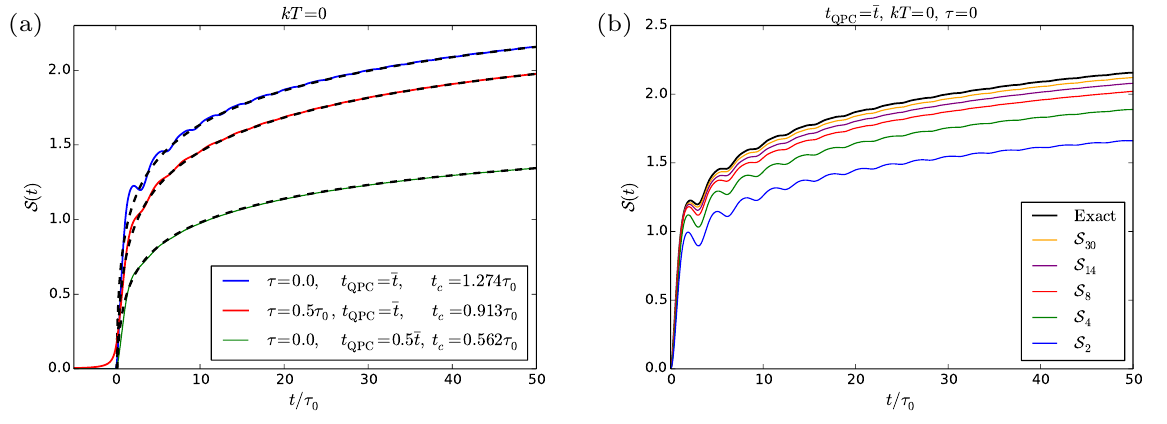}
	\caption{\label{Fig:EE_tau-t_Q}  (Color online) Entanglement entropy of the quantum switch. (a) Entanglement entropy for different opening times and transmissions of the QPC (full lines) together with the analytic expression (dashed lines) in Eq.~(\ref{eq:finite_size_prediction}). The effective central charge is $c_{\mathrm{eff}}\simeq1$ for full transmission and $c_{\mathrm{eff}}\simeq0.7$ for $t_{\mathrm{QPC}}=0.5\tb$ in good agreement with the prediction in Ref.~\onlinecite{Eisler2012}. There are $M=300$ sites in each lead. (b) Entanglement entropy for an abrupt opening. The black line shows the exact result while the colored lines are lower bounds obtained by including cumulants up to order $K$ in the series $\mathcal{S}_K$. The unit of time is $\tau_0=\hbar/\tb$.}
\end{figure*}

We are now ready to evaluate the entanglement entropy for a large tight-binding system with many particles. The example above makes it clear that a calculation of the entanglement entropy directly from its definition is very demanding when many particles are involved. However, as we will see, both the entanglement entropy and the FCS can be expressed in terms of a single-particle correlation matrix.\cite{Klich2009,Song2011,Song2012} These expressions also make it possible to relate the entanglement entropy to the cumulants of the FCS as shown in Refs.~\onlinecite{Klich2009,Song2011,Song2012}. In this section we define our tight-binding model together with the single-particle correlation matrix in terms of which we express the FCS and the entanglement entropy.

\subsection{Hamiltonian}

We work with non-interacting spinless fermions and can thus describe the system in Fig.~\ref{Fig:Fig1}a using a single-particle Hamiltonian. The Hamiltonian of the leads reads
\begin{equation}
\hatH_0=\sum_{\alpha=L,R}\hatH_\alpha,
\end{equation}
with lead $\alpha$ ($=L,R$) modeled as a tight-binding chain with nearest-neighbor hopping amplitude $\tb$,
\begin{equation}
\hatH_{\alpha}=-\tb\sum_{m=1}^{M_\alpha-1}\ket{m,\alpha}\bra{m+1,\alpha}+\hc,
\label{eq:H_alpha}
\end{equation}
and $M_\alpha$ sites labeled as $\{\ket{m,\alpha}\}$ with $m=1,\dots M_\alpha$. The total Hamiltonian reads
\begin{equation}
\hatH(t)=\hatH_0+\hatH'(t),
\end{equation}
where $\hatH'(t)$ is a time-dependent part that connects the two leads. Below, the time-dependent part describes the opening and closing of a QPC connecting the leads, as well as time-dependent pulses applied to the leads. Throughout the paper we take identical leads, $M_L=M_R=M$. Without a bias, the leads are at half filling, $N_0^L=N_0^R=M/2$, so that the chemical potentials are zero. Our calculations are typically performed with $M=200-300$ sites in each lead.

\subsection{Correlation matrix}

To evaluate the entanglement entropy and the FCS we need the correlation matrix $\hatM$ whose single-particle elements are the correlators\cite{Peschel2003}
\begin{equation}
[\hatM]_{i,j} =\ew{\hat{c}_{jR}^\dag \hat{c}_{iR}}
\label{eq:corr_matrix}
\end{equation}
with $i,j=1,\ldots, M_R$. If the two leads at the initial time $t=t_0$ are disconnected and in the uncorrelated product state $\hatrho=\hatrho_L\otimes\hatrho_R$ with $\hatrho_\alpha=e^{-\beta(\hatH_\alpha-\mu_\alpha \hatP_\alpha)}/\tr_\alpha[e^{-\beta(\hatH_\alpha-\mu_\alpha \hatP_\alpha)}]$, where $\beta=1/k_BT$ is the inverse electronic temperature and $\mu_\alpha$ is the chemical potential of lead $\alpha$, the correlation matrix reads\cite{Schonhammer2009,Abanov2009}
\begin{equation}
\hat{M}=\sqrt{\hatn_0}\hatU^\dag\hatP_R\hatU\sqrt{\hatn_0}+ (1-\hatn_0)\hatP_R.
\label{eq:M_T}
\end{equation}
Here $\hatP_\alpha$ is a projector onto the states in lead $\alpha$,
\begin{equation}
\hatP_\alpha=\sum_{m=1}^{M_\alpha}\ket{m,\alpha}\bra{m,\alpha},
\end{equation}
and
\begin{equation}
\hatn_0=\frac{1}{1+e^{\beta[\hatH(t_0)-\sum_{\alpha}\mu_\alpha \hatP_\alpha]}}
\label{eq:n_0}
\end{equation}
is the occupation-number operator at $t=t_0$. In addition, the time evolution operator reads
\begin{equation}
\hatU={\cal T}_t\exp\of{-\frac{i}{\hbar}\int_{t_0}^t dt'\hatH(t')},
\label{eq:TEO}
\end{equation}
where ${\cal T}_t$ is the time-ordering operator. In App.~\ref{App:TEO}, we describe how $\hatU$ is evaluated numerically.

If the leads are already connected at $t=t_0$ and the full system is in the correlated equilibrium state ($\mu_L=\mu_R=0$), $\hatrho=e^{-\beta\hatH(t_0)}/\tr[e^{-\beta\hatH(t_0)}]$, the correlation matrix reads
\begin{equation}
\hatM=\hatP_R\hatU\hatn_0\hatU^\dag\hatP_R,
\label{eq:M_0}
\end{equation}
as shown in Ref.~\onlinecite{Klich2009}.

\subsection{Full counting statistics}

The FCS of transferred charge can be expressed in terms of the correlation matrix as\cite{Abanov2009}
\begin{equation}
\chi(\lambda)=\det[1+(e^{i\lambda}-1)\hat M]e^{-i\lambda q},
\label{eq:MGF}
\end{equation}
with $q=\tr[\hatM]|_{t=t_0}$, such that the first cumulant $C_1(t_0)$ initially is zero. Using the relation $\ln(\det[\hat A])=\tr[\ln \hat  A]$ for a matrix $\hat  A$, the cumulant of order $m\geq2$ follows as
\begin{equation}
\label{eq:cumulantsinM}
C_m=\sum_{k=1}^m c_k^{(m)} \tr[\hatM^k],
\end{equation}
having defined the coefficients
\begin{equation}
c_k^{(m)}=\sum_{l=1}^k (-1)^{(l-1)}\binom{k-1}{l-1}l^{(m-1)}.
\end{equation}
Combined with the correlation matrix in Eq.~(\ref{eq:corr_matrix}), these expressions allow us to calculate the cumulants of the charge transport as functions of time.

\subsection{Entanglement entropy}

The entanglement entropy can be obtained from the correlation matrix using the expression\cite{Klich2009}
\begin{equation}
\EE=-\tr\left[\hatM\ln\hatM-(1-\hatM)\ln(1-\hatM)\right].
\label{eq:EEinM}
\end{equation}
We subtract any initial entropy and always show the increase of the entanglement entropy $\EE(t)-\EE(t_0)$. The R\'enyi entropies can similarly be written as\cite{Song2012}
\begin{equation}
\EE^{(\nu)}=\frac1{1-\nu}
\tr\left[\ln\left\{\hatM^\nu+(1-\hatM)^\nu\right\}\right].
\label{eq:REinM}
\end{equation}
These equations make it possible to relate the entanglement and R\'enyi entropies to the cumulants of the FCS. Specifically, as shown in Refs.~\onlinecite{,Song2011,Song2012}, by expanding Eqs.~(\ref{eq:EEinM}) and (\ref{eq:REinM}) in powers of $\hatM$ combined with Eq.~(\ref{eq:cumulantsinM}), one arrives at the series in Eqs.~(\ref{eq:EEcumulants}) and (\ref{eq:REcumulants}) for the entropies in terms of the cumulants.

\begin{figure*}
	\includegraphics[width=0.95\textwidth]{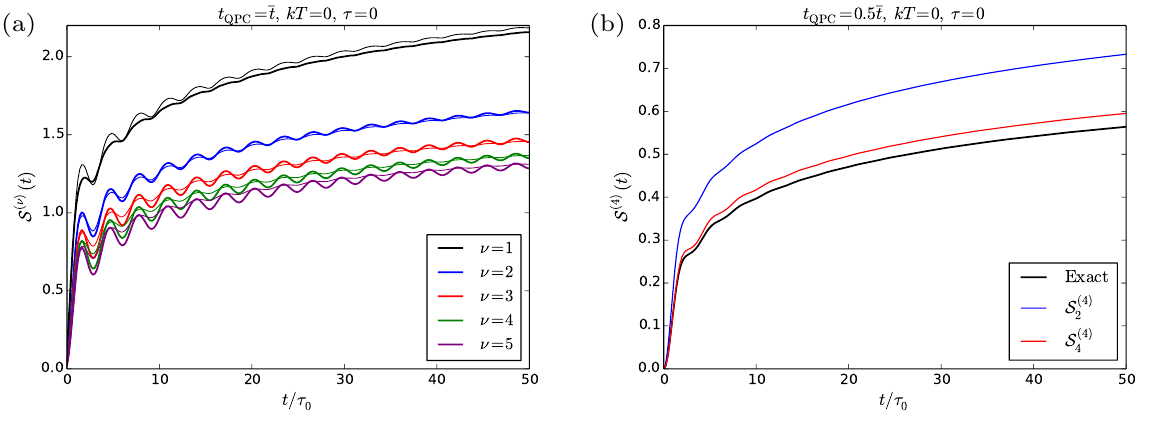}
	\caption{\label{Fig:RE_QPC} (Color online) R\'enyi entropies of the quantum switch. (a) R\'enyi entropies for a fully transmitting QPC opened abruptly at $t=0$. The thin lines indicate the gaussian approximation where only the second cumulant is taken into account. (b) R\'enyi entropy of order $\nu=4$ for a QPC opened abruptly at $t=0$ to $\tq=0.5\tb$. The exact result (black line) is shown together with the series $\mathcal{S}^{(4)}_K$ with cumulants up to order $K$ (colored lines). The unit of time is $\tau_0=\hbar/\tb$.}
\end{figure*}

\section{Quantum switch}
\label{sec:Switch}

We consider the quantum switch depicted in Fig.~\ref{Fig:Fig1}.\cite{Klich2009} It consists of a QPC with a transmission that is varied in time. The time-dependent Hamiltonian reads
\begin{equation}
\hatH'(t)=-f(t)\left(\tq\ket{1,L}\bra{1,R}+\hc\right),
\label{eq:H_tot_QPC}
\end{equation}
with $f(t)$ controlling the transmission of the QPC and $\tq\leq\tb$. We first consider the situation where the QPC initially is closed. We then open it by choosing
\begin{equation}
f(t)=\frac12+\frac1{\pi}\arctan\of{\frac{t}{\tau}}.
\end{equation}
Here $\tau$ is the opening time of the QPC with $\tau=0$ corresponding to an abrupt opening. We initialize the disconnected leads at $t_0\ll -\tau$, where $f(t_0)\approx0$. For an abrupt opening, the setup bears some similarities with the quantum Ising chain investigated in Ref.~\onlinecite{Igloi2009}.

Figure~\ref{Fig:EE_tau-t_Q}a shows the time evolution of the entanglement entropy upon opening the QPC at $t=0$. The temperature is zero and no bias voltage is applied between the leads. For finite-size leads, the entanglement entropy is expected to follow the prediction\cite{Stephan2011,Eisler2012}
\begin{equation}
\EE=(c_{\mathrm{eff}}/3)\ln|(t_M/t_c)\sin(t/t_M)|,
\label{eq:finite_size_prediction}
\end{equation}
where the time scale $t_M=2M/(\pi v_F)$ is determined by the number of sites $M$ in each lead, the short-time cutoff $t_c$ is on the order of the opening time $\tau$,\cite{Klich2009,Song2011,Song2012} and $c_{\mathrm{eff}}$ is the (effective) central charge of the conformal field theory, depending on the transmission of the QPC.\cite{Eisler2012} For a fully transmitting QPC connecting infinitely long leads, $M\rightarrow\infty$, this expression reduces to the seminal result\cite{Holzhey1994,Vidal2003,Calabrese2004}
\begin{equation}
\EE=(1/3)\ln(t/t_c)
\label{eq:CFTprediction}
\end{equation}
from conformal field theory. Our results clearly follow Eq.~(\ref{eq:finite_size_prediction}) with small oscillations on top of the logarithmic growth if the QPC is abruptly opened. The frequency of the oscillations is given by the distance of the chemical potential to the nearest band edge, $\omega_0=2\tb-|\mu|$, as we have found by systematically varying the occupation of the leads.  In addition, the oscillations are smeared out when the opening of the QPC is smooth.

In Fig.~\ref{Fig:EE_tau-t_Q}b we turn to the series expansion of the entanglement entropy in terms of the measurable cumulants of the FCS. The figure demonstrates how the exact result is approached as more cumulants are included. In this example, cumulants of very high orders ($K\simeq 30$) are needed for the series to converge. However, already with the second cumulant ($K=2$) only, the series provides a good approximation of the exact entanglement entropy.

Figure~\ref{Fig:RE_QPC}a shows the time evolution of the R\'enyi entropies up to order $\nu=5$. In this case, we compare the exact results with the gaussian approximations in Eqs.~(\ref{eq:EE_Gauss}) and (\ref{eq:RE_Gauss}). For a fully open QPC ($\tq=\tb$), the gaussian approximation works well. Thus, from a measurement of the second cumulant only, one obtains a good approximation of the entanglement and the R\'enyi entropies. This only holds true for a fully transmitting QPC as illustrated in Fig.~\ref{Fig:RE_QPC}b, showing the series for $\mathcal{S}^{(4)}$ in terms of a finite number of cumulants for a QPC with non-unity transmission. In this case, it is necessary to go beyond the second cumulant before convergence is reached. However, already when the fourth cumulant is included, the series provides a good approximation of the exact R\'enyi entropy.

\begin{figure*}
	\includegraphics[width=0.95\textwidth]{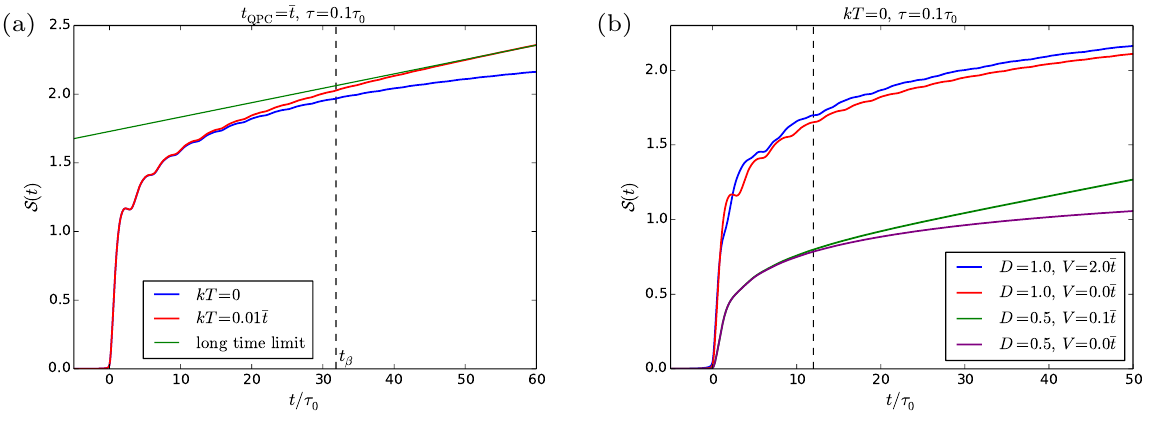}
    \caption{\label{Fig:EE_finite_T}  (Color online) Entanglement entropy at finite temperatures and bias. (a) Entanglement entropy for a QPC with a finite electronic temperature (red). Around $t_\beta=\hbar/(\pi kT)$, the growth of the entanglement entropy changes from logarithmic to being linear in time. The zero temperature result (blue) is shown for comparison together with the linear long-time asymptotics (green). (b) Entanglement entropy for a finite bias $V$ at zero temperature. For a fully open QPC, the entanglement entropy still grows logarithmically with time. For non-unit transmission, the entanglement entropy becomes linear in time for $t\gtrsim h/V$ (dashed vertical line).  The unit of time is $\tau_0=\hbar/\tb$.}
\end{figure*}

Next, we consider the influence of a finite electronic temperature as well as a finite bias voltage between the leads. The results shown so far were obtained at zero temperature, where the full system is in a pure state and the growth of the entanglement entropy is only due to quantum fluctuations between the leads. With a finite electronic temperature, thermal fluctuations come into play together with shot noise due to the applied voltage. This is illustrated in Fig.~\ref{Fig:EE_finite_T}a, where we show the time-dependent entanglement entropy for a finite electronic temperature. The generalization of the prediction in Eq.~(\ref{eq:CFTprediction}) then reads\cite{Song2012,Swingle2013}
\begin{equation}
\EE=\frac{1}{3}\ln\left\{\frac{t_\beta}{t_c}\sinh\left(\frac{t}{t_\beta}\right)\right\},
\label{eq:CFTprediction_temp}
\end{equation}
where $t_\beta=\hbar/(\pi kT)$ is a thermal time scale. At short times, $t\ll t_\beta$, the logarithmic behavior from Eq.~(\ref{eq:CFTprediction}) persists. In contrast, at long times, $t\gg t_\beta$, thermal fluctuations in the Fermi seas cause the entanglement entropy to grow linearly in time. This crossover is captured by Fig.~\ref{Fig:EE_finite_T}a, showing the entanglement entropy with and without a finite electronic temperature.

In Fig.~\ref{Fig:EE_finite_T}b, we consider a quantum switch with a finite bias voltage between the leads. We take the left and right leads as the source and drain electrodes, respectively, so that $V=\mu_L-\mu_R$ is the potential difference between the leads. We mimic a finite bias by choosing the initial particle numbers of the leads differently, giving an approximately constant current during a certain time window.\cite{Schonhammer2007,Thomas2014} Specifically, we choose the initial occupations as $N_0^{L/R}=(M\pm\Delta N)/2$, where $\Delta N$ is the surplus of particles in the source electrode. We then have $\mu_L=-\mu_R$ together with the approximate potential drop
\begin{equation}
V\simeq4\tb\sin\of{\frac\pi2\frac {\Delta N}{M}}
\end{equation}
for large tight-binding leads, $M\gg1$.

Figure~\ref{Fig:EE_finite_T}b shows that the entanglement entropy for a fully transmitting QPC essentially grows logarithmically with time just as in the unbiased case. In contrast, for a QPC with a transmission below unity, electrons in the transport window may reflect back on the QPC, generating shot noise. These transport processes can be considered as binomial events, where each electron with probability $D$ is transmitted through the QPC and with probability $1-D$ is reflected. For such binomial processes, the entanglement entropy is expected to become linear in time at long times following the expression\cite{Beenakker2006}
\begin{equation}
\EE(t)=-\frac{t}{\bar{\tau}}\left[D\ln D+(1-D)\ln(1-D)\right],
\label{eq:EE_finite_V}
\end{equation}
where $\bar{\tau}=h/eV$ is the mean waiting time between the incoming electrons,\cite{Albert2012} and the ratio $t/\bar{\tau}$ yields the number of transmission attempts after the QPC has been opened. The crossover to the linear behavior is seen in Fig.~\ref{Fig:EE_finite_T}b for a biased QPC with a non-unity transmission.

\begin{figure}
    \includegraphics[width=0.95\columnwidth]{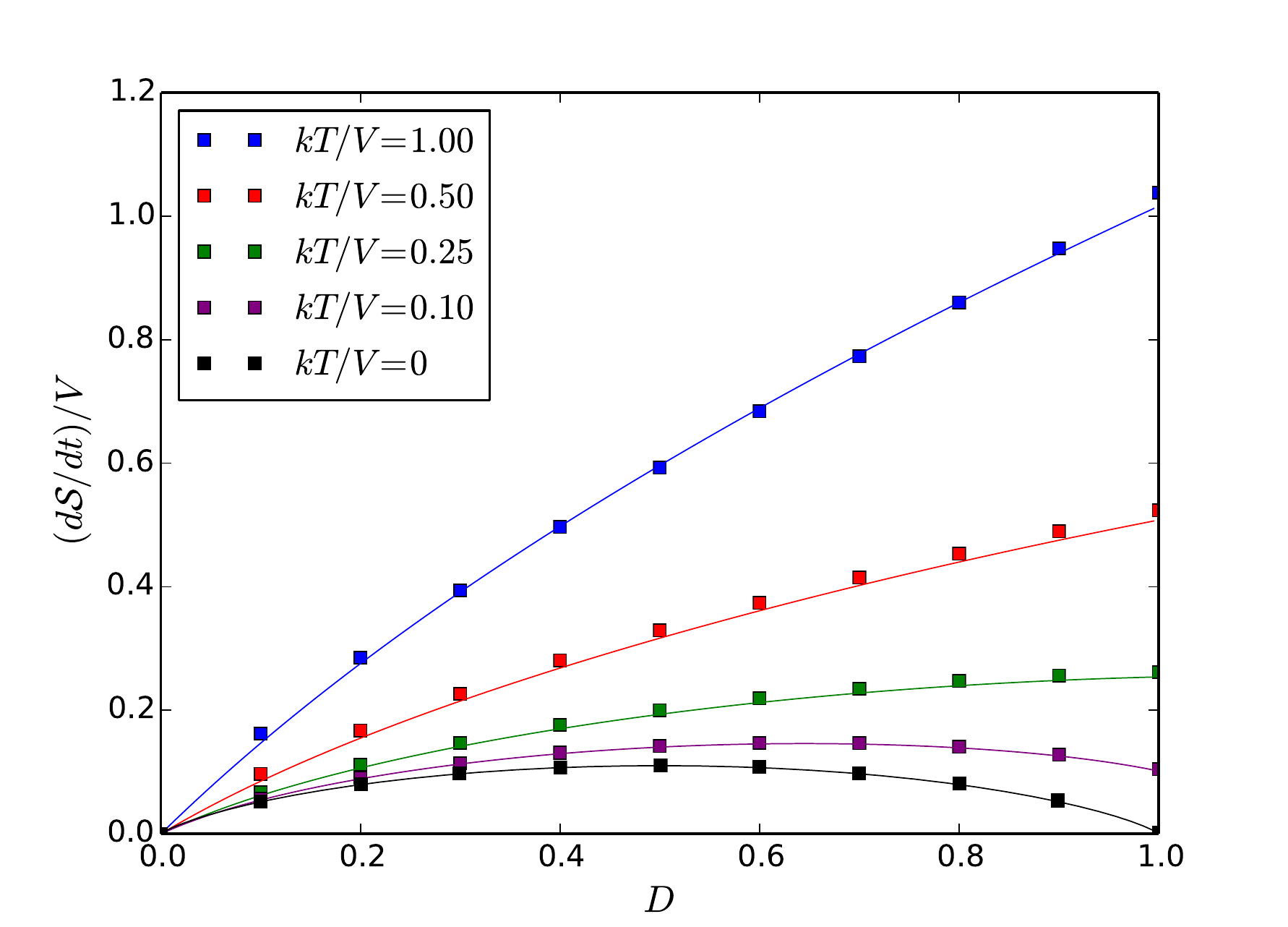}
	\caption{\label{Fig:EE_finite_V-D=1}  (Color online) Growth of the entanglement entropy at long times. We show the derivative of the entanglement entropy $d\EE/dt$ with respect to time at long times, $t\gg h/V$. Results are shown as functions of the transmission $D$ of the QPC for different ratios of the temperature over the bias. The squares are the results of our numerically exact calculations, whereas the solid lines are obtained from Eq.~(\ref{eq:CGF_eV}).}
\end{figure}

To understand the combined effect of a finite voltage bias and a finite electronic temperature, we consider in Fig.~\ref{Fig:EE_finite_V-D=1} the derivative of the entanglement entropy with respect to time for different temperatures and voltages as well as different transmissions of the QPC. Neglecting the quantum noise at zero bias and zero temperature and assuming an energy-independent transmission $D$, the logarithm of the moment generating function reads\cite{Levitov1996}
\begin{equation}
\label{eq:CGF_eV}
\ln\chi(\lambda)=-\frac{tkT}{h}u_+u_-,
\end{equation}
where
\begin{equation}
u_\pm=v\pm\mathrm{Arcosh}\left[D\cosh(v+i\lambda)+(1-D)\cosh(v)\right],
\nonumber
\end{equation}
and $v=V/(2kT)$ is the ratio between the potential difference and the temperature. For our tight-binding chain, the transmission probability is energy-dependent and reads\cite{Thomas2014}
\begin{equation}
{\cal T}(\eps)=\frac{\theta^2(4-(\eps/\tb)^2)}{1+\theta^2(2-(\eps/\tb)^2)+\theta^4},
\label{eq:T_eps}
\end{equation}
with $\theta=\tq/\tb$. The transmission is perfect ${\cal T}(\eps)=1$ for $\tq=\tb$. To make a connection with Eq.~(\ref{eq:CGF_eV}), we average the energy-dependent transmission over the bias window and take
\begin{equation}
D=\overline{{\cal T}(\eps)}=\frac1{V}\int_{-V/2}^{V/2}d\eps\,{\cal T}(\eps),
\label{eq:D_def}
\end{equation}
where $V/2=\mu_L=-\mu_R$ is half the symmetrically applied bias. The averaged transmission then becomes
\begin{equation}
D=1-\frac{2\tb}{V}\frac{(1-\theta^2)^2}{\theta(1+\theta^2)}\mathrm{artanh}\left[\frac {V}{2\tb}\frac\theta{1+\theta^2}\right].
\label{eq:D_Fermi}
\end{equation}
Using this expression for the transmission probability, our results for the entanglement entropy in Fig.~\ref{Fig:EE_finite_V-D=1} are in excellent agreement with predictions based on Eq.~(\ref{eq:CGF_eV}). We note that our results for low temperatures can also been obtained by inserting the energy-dependent
transmission from Eq.~(\ref{eq:T_eps}) into Eq.~(\ref{eq:EE_finite_V}) and then average over the energy, as we have checked.\cite{footnote1}

From the findings above we see that the entanglement entropy increases due to three types of processes: quantum noise at zero bias and zero temperature, which gives rise to the logarithmic behavior in Eq.~(\ref{eq:CFTprediction}), together with thermal and shot noise fluctuations which cause a linear increase with time. In an experiment, the shot noise contribution can be suppressed simply by not applying a bias. However, thermal fluctuations, which dominate over the quantum noise at long times, will always be present. Thus, to access the logarithmic short-time behavior due to quantum noise, it has been suggested to open and close the QPC in a periodic manner.\cite{Klich2009}

\begin{figure}
	\includegraphics[width=0.95\columnwidth]{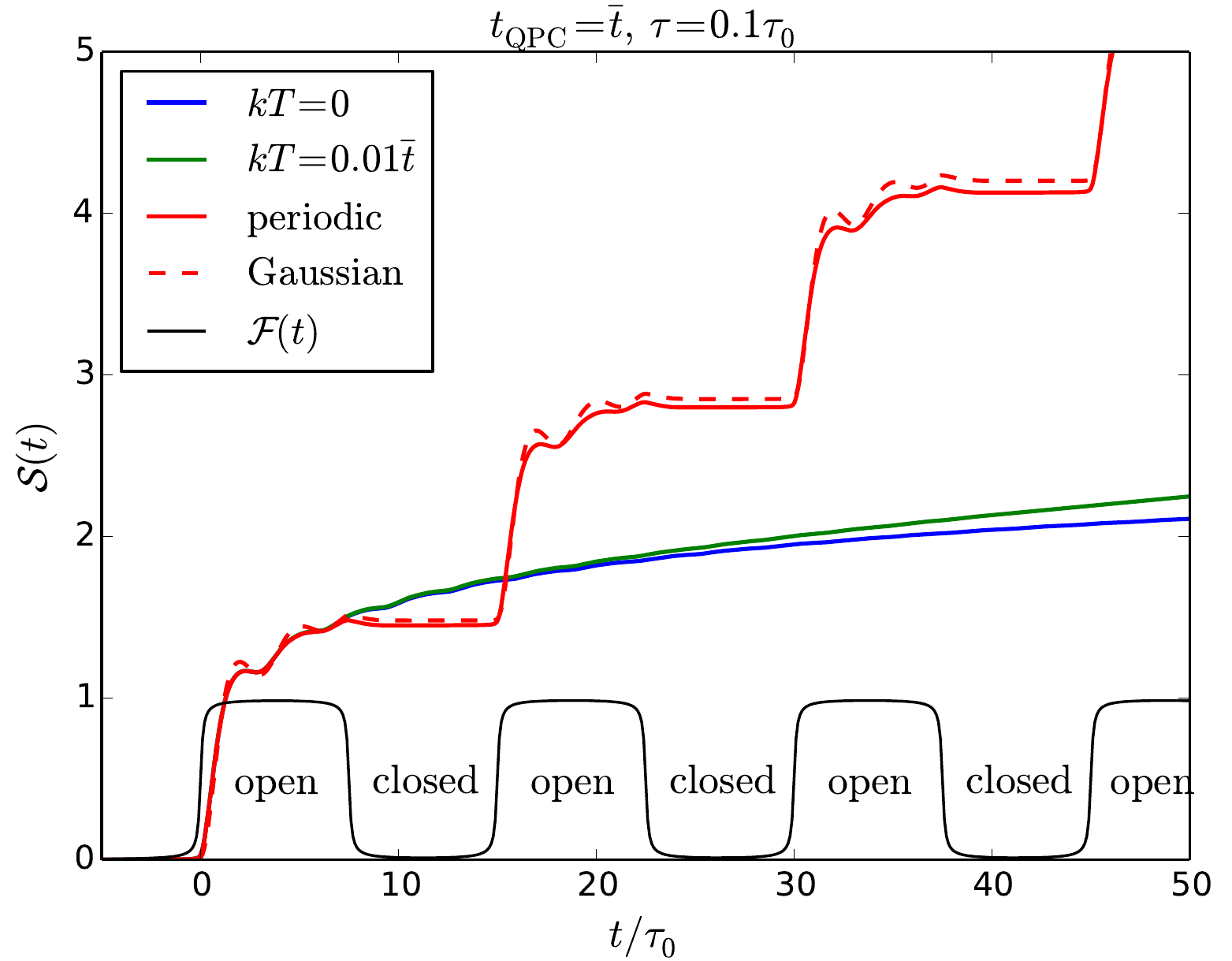}
	\caption{\label{Fig:EE_finite_T_periodic} (Color online) Driven quantum switch. The transmission of the QPC is controlled by the function $\mathcal{F}(t)$ in Eq.~(\ref{eq:periodic_opening}). During each period of duration $\mathcal{T}=15\tau_0$ with $\tau_0=\hbar/\tb$, the QPC is opened for the time $w=7.5\tau_0$. Since this is shorter than the thermal time scale $t_\beta=\hbar/(\pi kT)\simeq 32\tau_0$, the logarithmic growth dominates within each period. With $T=10$ mK, we have $t_\beta\simeq 0.24$ ns which should be reachable with current technology. The dashed line shows the gaussian approximation of the entropy.}
\end{figure}

To describe the periodic opening and closing of the QPC, we replace $f(t)$ in Eq.~(\ref{eq:H_tot_QPC}) by the periodic function
\begin{equation}
\mathcal{F}(t)=\sum_{n=0}^\infty\left[f(t-n\mathcal{T})-f(t-w-n\mathcal{T})\right].
\label{eq:periodic_opening}
\end{equation}
Here $\mathcal{T}$ is the period of the driving and $w$ is the length of the time windom during which the QPC is open. Choosing the time window to be shorter than the thermal time scale, $w<t_\beta$, we expect to see a recurrence of the logarithmic behavior after each opening of the QPC. This is confirmed by our calculations in Fig.~\ref{Fig:EE_finite_T_periodic}. Importantly, the exact results for the entanglement entropy are very well captured by the gaussian approximation which only includes the second cumulant. Moreover, the second cumulant of the full counting statistics can be related to the current noise, thereby paving the way for an experimental verification of predictions from conformal field theory in a coherent electronic conductor.

\section{Levitons}
\label{sec:Levitons}

\begin{figure*}
\includegraphics[width=2\columnwidth]{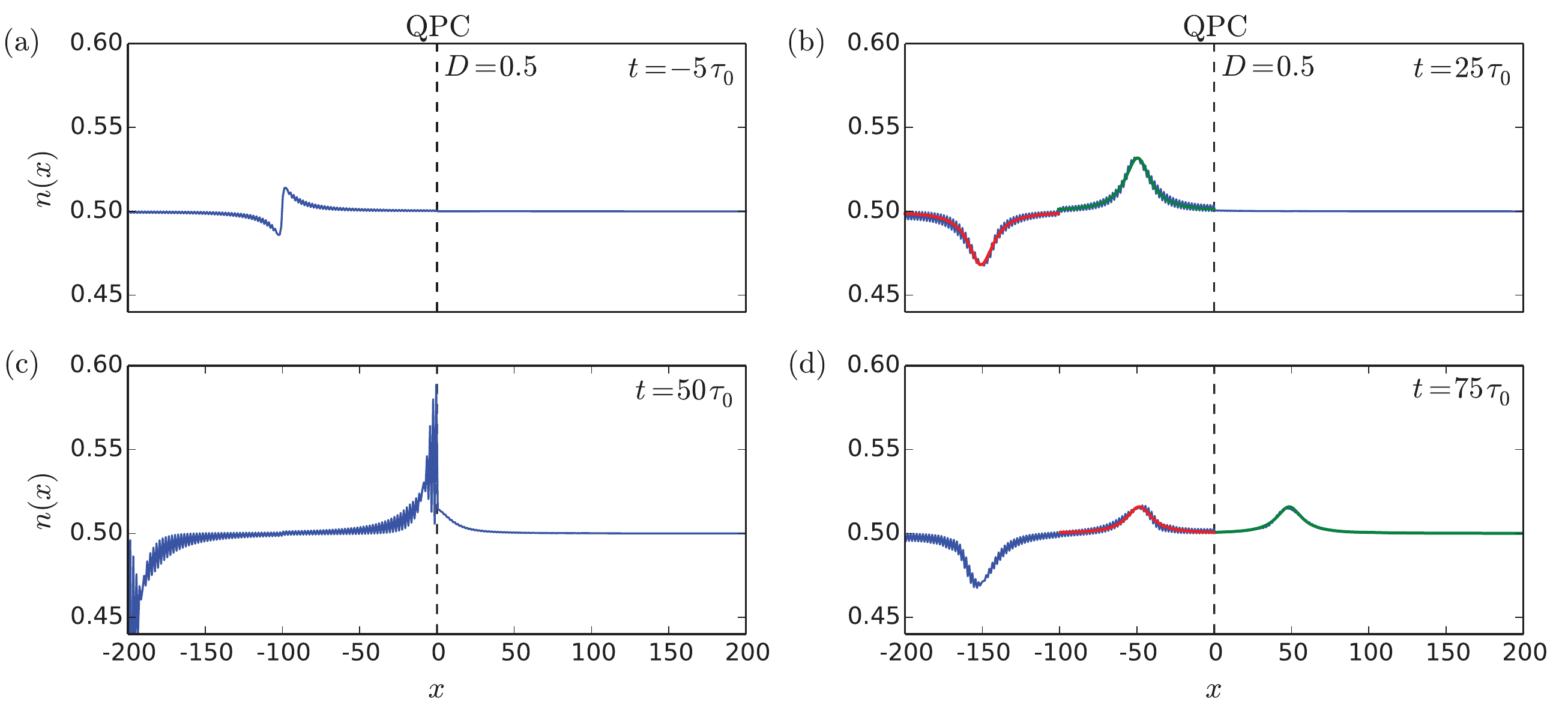}
\caption{(Color online) Creation of levitons. We show the time-evolution of the particle density $n(x)$ along a tight-binding chain with a QPC placed at $x=0$. A leviton and an anti-leviton are created by modulating the phase of the hopping amplitude in the middle of the left lead according to Eq.~(\ref{eq:phi_Lev}). Panel (a) shows the leviton and the anti-leviton emerging out of the Fermi sea.  Panel (b) shows the two quasi-particles propagating in opposite directions. The red and green lines indicate their lorentzian density profiles. In panel (c), the leviton scatters on the QPC tuned to half transmission, while the anti-leviton is reflected at the end of the chain. Panel (d) shows the transmitted and reflected parts of the leviton, which again have lorentzian density profiles.  The unit of time is $\tau_0=\hbar/\tb$.}
\label{Fig:Leviton_desity}
\end{figure*}

Motivated by recent experiments,\cite{Dubois2013nature,Jullien2014} we now consider time-dependent excitations of the Fermi sea. In the experiment by Dubois and co-workers,\cite{Dubois2013nature} lorentzian-shaped voltage pulses applied to an ohmic contact led to the creation of clean single particle excitations (levitons), following a theoretical proposal by Levitov and co-workers.\cite{Levitov1993,Levitov1996,Keeling2006} Similar excitations can be generated by applying a slow linear drive to a quantum capacitor.\cite{Keeling2008,Feve2007} A third strategy, which applies to our tight-binding chain, is to modulate the phase of the tunneling amplitude between neighboring sites as we show in App.~\ref{App:Levs_at_QPC}. In this case, the time-dependent part of the Hamiltonian reads
\begin{equation}
\label{eq:H_T_Lev}
\begin{split}
\hatH'(t)=&-\tq\ket{1,L}\bra{1,R}+\hc\\
&-\tb\sum_{\{j,\alpha\}}e^{i\phi_{j\alpha}(t)}\ket{j,\alpha}\bra{j-1,\alpha}+\hc,
\end{split}
\end{equation}
where the first line describes the (static) coupling of the leads due to the QPC and the terms on the second line create levitons at site $j$ in lead $\alpha=L,R$. (The sum runs over the sites $\{j,\alpha\}$, where we wish to create levitons). As we show in App.~\ref{App:Levs_at_QPC}, the phase of the tunneling amplitude should be chosen as
\begin{equation}
\label{eq:phi_Lev}
\phi_{j\alpha}(t)=2\arctan\of{\frac{t-t_{j\alpha}}\tau}+\pi,
\end{equation}
where $t_{j\alpha}$ is the emission time and $\tau$ determines the width of the lorentzian wave packet that is produced. With this phase, a right-moving leviton is created together with a left-moving anti-leviton (a hole), but without additional electron-hole pairs. If the sign of the phase is changed, the quasi-particles move in the opposite directions.

Figure~\ref{Fig:Leviton_desity} shows the particle density $n(x)$ along the chain at different times. The QPC is positioned at $x=0$ with $x<0$ ($x>0$) corresponding to the left (right) lead. The system is initialized in equilibrium at half filling at the time $t_0\ll \min\{t_{j\alpha}\}$, long before any excitation is applied. As the pulse is applied, a leviton and an anti-leviton are generated in the left lead as seen in panel (a). The excitations propagate in opposite directions with the Fermi velocity $v_F=2\tb/\hbar$. In a continuum description, the wave function of the leviton reads\cite{Keeling2006}
\begin{equation}
\psi_{\pm}(x,t)=\frac{i\sqrt{v_F\tau/\pi}}{x-x_0\pm v_F(t-t_e)+iv_F\tau},
\label{eq:levitonPsi}
\end{equation}
where $x_0$ is the position at which the leviton is created, and the sign corresponds to a leviton moving to the left ($-$) or to the right ($+$). With this wave function, the corresponding particle density $n(x)$ becomes lorentzian, which agrees well with our results in Fig.~\ref{Fig:Leviton_desity}.

\begin{figure}
\includegraphics[width=0.95\columnwidth]{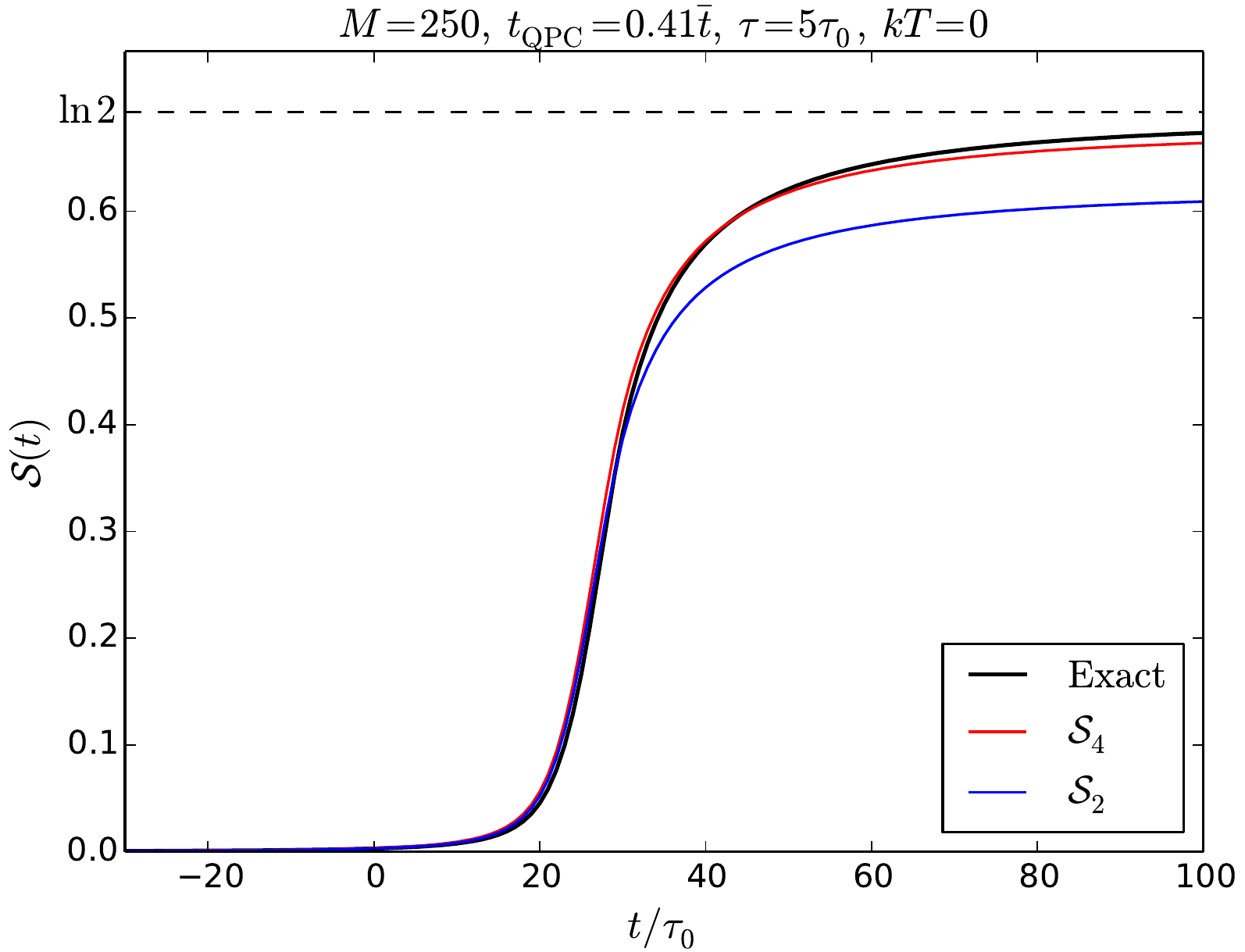}
\caption{(Color online) Entanglement entropy for a single leviton impinging on a QPC. The leviton is created at time $t_{j\alpha}=0$ at site $j=50$ in the left lead ($\alpha=L$). The entanglement entropy increases nearly by $\ln 2$. The colored lines show the cumulant series with an increasing cut-off $K$. The unit of time is $\tau_0=\hbar/\tb$.}
\label{Fig:EE_Leviton}
\end{figure}

As the leviton scatters on the QPC, it is partitioned into a transmitted and a reflected wave packet. The QPC is tuned to half transmission, such that transmission and reflection occur with equal weight. In Fig.~\ref{Fig:EE_Leviton} we show the time evolution of the entanglement entropy during the scattering process. Only the first few cumulants are needed to obtain the entanglement entropy from the series in Eq.~(\ref{eq:EEcumulants}) and already with $K=6$ the agreement with the full result is very good. For a binomial process with fifty percent success probability, the entropy should increase by $\ln 2$. Our results are close to this value, although slightly lower due to the exponential distribution of the leviton in the energy domain,\cite{Keeling2006} $\psi(\eps)\propto e^{-\eps\tau}\Theta(\eps)$. As a result, different components of the wave packet are scattered with different transmission amplitudes, leading to a smaller increase in the entropy. We note that the entanglement generated here comes from a superposition of states with different particle numbers in each lead and thus may not be accessible.

\begin{figure}
\includegraphics[width=0.95\columnwidth]{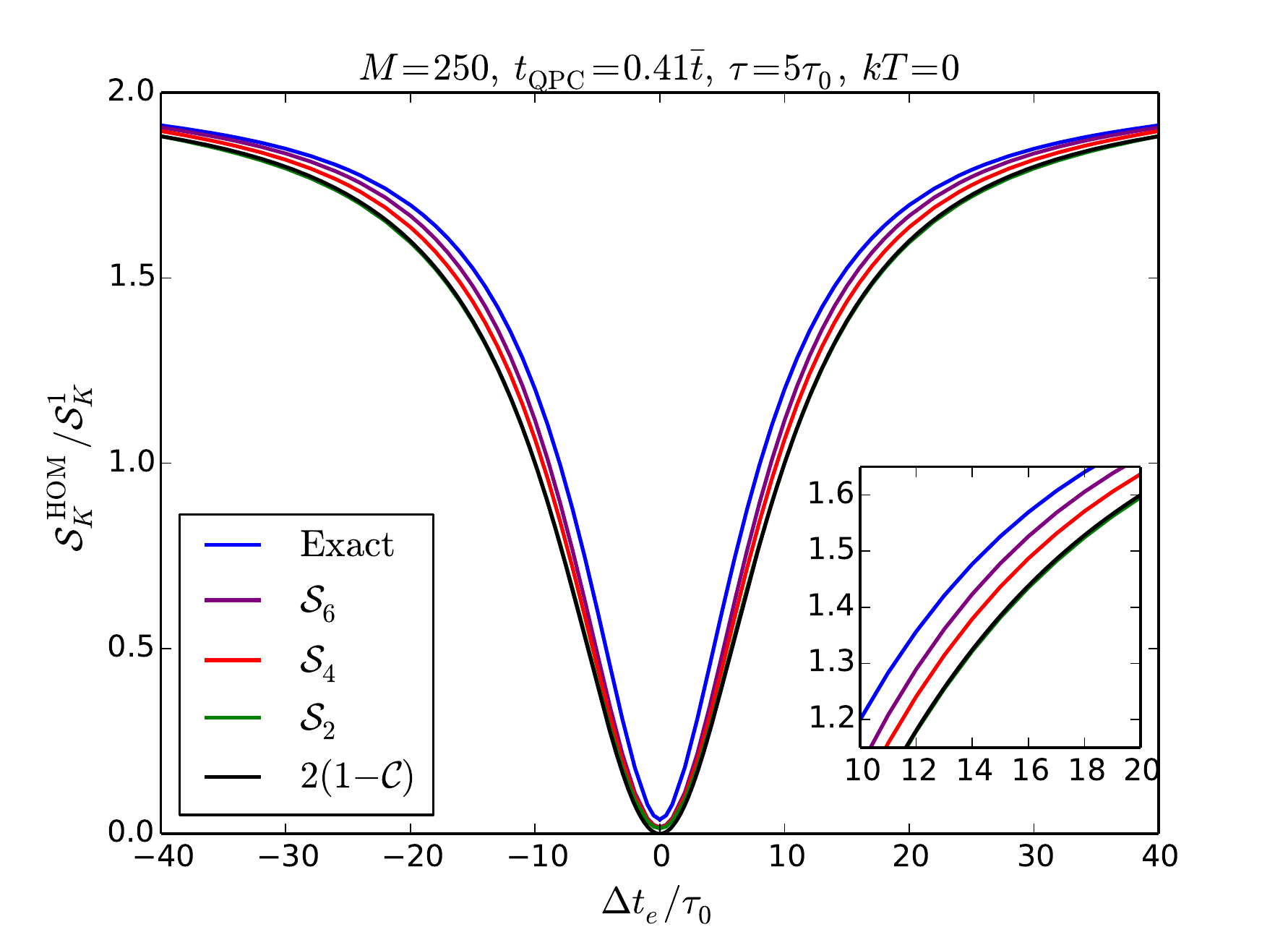}
\caption{(Color online) Entanglement entropy in a fermionic Hong-Ou-Mandel experiment with levitons. The entanglement entropy exhibits a Hong-Ou-Mandel-like suppression as a function of the  time delay $\Delta t_e$ between the arrival times at the QPC. We show the exact result for the entanglement entropy together with the cumulant series using an increasing cut-off $K$. The black line indicates the expected increase of the second cumulant according to Eq.~(\ref{eq:S_HOM}). The inset focuses on the differences between the curves. The unit of time is $\tau_0=\hbar/\tb$.}
\label{Fig:EE_Levitons}
\end{figure}

Following the work of Dubois and co-workers,\cite{Dubois2013nature} we now consider the situation where one leviton is created in each lead and brought into collision at the QPC in a fermionic Hong-Ou-Mandel experiment.\cite{Hong1987,Bocquillon2013} In Fig.~\ref{Fig:EE_Levitons} we show the entanglement entropy generated by interfering two levitons on the QPC as a function of the time delay $\Delta t_e$ between the arrival times at the QPC. The results are divided by the entanglement entropy generated by scattering just a single leviton on the QPC. Together with the full result, we show the entanglement entropy obtained using the cumulant series with an increasing cut-off.

If the levitons arrive simultaneously, they anti-bunch such that one leviton leaves the QPC in each direction after the scattering event. In this case, there are essentially no charge fluctuations and almost no entanglement entropy is generated. In contrast, for large time differences, $|\Delta t_e|\gg\tau$, the levitons scatter independently of each other and the entanglement entropy equals twice the entropy generated by a single scattering event. The final state is a coherent superposition of states with zero, one, and two levitons in one lead. The state with one leviton in each lead is time-bin entangled with finite accessible entanglement entropy.

To understand the shape of the curve in between these limiting situations, we consider the increase of the second cumulant following a Hong-Ou-Mandel experiment with levitons. It can be written as\cite{Dubois2013PRB} (see also Ref.~\onlinecite{Hofer2013})
\begin{equation}
C_2^\textrm{HOM}=2C_2^{1}(1-{\cal C}),
\label{eq:S_HOM}
\end{equation}
where $C_2^{1}$ is the increase for a single leviton and
\begin{equation}
{\cal C}=|\ew{\psi_{-}|\psi_+}|^2=\frac1{1+\of{\frac{\Delta t_e}{2\tau}}^2}
\end{equation}
is the overlap of the leviton wave functions, taking $x_0\mp v_F t=0$ in Eq.~(\ref{eq:levitonPsi}), depending on $\Delta t_e$.

This expression is in good agreement with our results for the second cumulant and it essentially determines the shape of the entanglement entropy as a function of $\Delta t_e$ with only minor corrections due to higher cumulants.

\section{Conclusions}
\label{sec:Conclusion}

We have investigated the entanglement and the R\'enyi entropies for two Fermi seas connected via a barrier with a time-dependent transmission. This system can be implemented in mesoscopic physics by connecting two electrodes via a QPC, or in an optical lattice with cold fermionic atoms. Using exact expressions for the entropies in terms of the cumulants of the FCS, we have shown how the entanglement and the R\'enyi entropies in a quantum many-body system can be deduced from measurements of the charge fluctuations between the reservoirs. In particular, for a quantum switch operated under suitable experimental conditions, a logarithmic growth of the entanglement entropy, as predicted by conformal field theory, can be inferred from only the first few cumulants. Motivated by recent experiments, we have evaluated the entanglement entropy generated by partitioning clean single-particle excitations (levitons) on a QPC as well as by interfering two levitons on the QPC, tuned to half transmission. In this case, we identify a Hong-Ou-Mandel-like suppression of the entanglement entropy as a function of the difference of arrival times at the QPC.

We hope our work may stimulate further theoretical efforts to understand entanglement entropy in condensed matter physics, for instance in relation to the flow of R\'enyi entropies in quantum heat engines.\cite{Nazarov2011,Ansari2014} In addition, the results presented here may serve as a guideline for future experiments aimed at measuring the entanglement entropy in a solid-state system.

\section*{Acknowledgements}

We thank David Dasenbrook, Viktor Eisler, and Patrick P.~Hofer for useful discussions. CF is affiliated with Centre for Quantum Engineering at Aalto University. The work was supported by the Swiss National Science Foundation.

\appendix

\section{Time evolution operator}
\label{App:TEO}

The time evolution operator $\hatU$ in Eq.~(\ref{eq:TEO}) can be evaluated using a Crank-Nicolson scheme.\cite{Press2007} By discretizing time in small steps of length $\delta t\ll\hbar/\tb$, we can write
\begin{equation}
\hatU(t+\delta t)\simeq e^{-i\hatH(t+\delta t/2)\delta t/\hbar}\hatU(t),
\end{equation}
assuming that the Hamiltonian,  here evaluated at the center of the interval, is roughly constant during the time step. We rewrite this approximation as
\begin{equation}
e^{i\hatH(t+\delta t/2)\delta t/2\hbar}\hatU(t+\delta t)\simeq e^{-i\hatH(t+\delta t/2)\delta t/2\hbar}\hatU(t)
\end{equation}
and expand the exponentials on each side to first order in $\delta t$. We then find the expression
\begin{equation}
\hatU(t+\delta t)\simeq \frac{2\hbar-i\hatH(t+\delta t/2) \delta t}{2\hbar+i\hatH(t+\delta t/2) \delta t}\hatU(t),
\end{equation}
which allows us to determine the time evolution operator iteratively in each time step, starting from the initial condition $\hatU(t_0)=\hat{1}$.

\section{Zero-frequency noise}
\label{App:Noise_and_C_2}

Here we show that the zero-frequency noise for a periodic process is given by the increase of the second cumulant per period following Ref.~\onlinecite{Camalet2004}. We first introduce the current-current correlation function
\begin{equation}
c(t,t')=\frac{1}{2}\langle\{\delta\hat I(t),\delta\hat I(t')\}\rangle,
\end{equation}
where the curly brackets denote the anti-commutator and $\delta\hat I(t)= \hat I(t)-\langle \hat I(t)\rangle$. For a periodic process, the correlation function shares the periodicity of the process, $c(t+\mathcal{T},t'+\mathcal{T})=c(t,t')$, where $\mathcal{T}$ is the period. The zero-frequency noise is then
\begin{equation}
S(0)=\frac{1}{\mathcal{T}}\int_0^\mathcal{T} dt\int_{-\infty}^{\infty}d\tau c(t,t-\tau).
\end{equation}

To relate the zero-frequency noise to the second cumulant of the FCS, we consider the charge $\hat Q(t)=\int_{t_0}^t dt' \hat I(t')$ accumulated in one of the leads during the time interval~$[t_0,t]$. The second cumulant of the charge fluctuation can then be written as
\begin{equation}
C_2(t)=\langle[\delta\hat Q(t)]^2\rangle=\langle\hat Q^2(t)\rangle-\langle\hat Q(t)\rangle^2.
\end{equation}
with $\delta\hat Q(t)=\hat Q(t)-\ew{\hat Q(t)}$. Differentiating this expression with respect to time, we find
\begin{equation}
\begin{split}
\frac d{dt}C_2(t)&=\langle\{\delta\hat I(t),\delta\hat Q(t)\}\rangle\\
&=\int_{t_0}^{t} dt'\langle\{\delta\hat I(t),\delta\hat I(t')\}\rangle\\
&=2\int_0^{t-t_0}d\tau c(t,t-\tau).
\end{split}
\end{equation}
Finally, by averaging over one period and taking the limit $t_0\to-\infty$, we find
\begin{equation}
\frac{1}{\mathcal{T}}\int_0^\mathcal{T}dt \frac d{dt}C_2(t)=\frac{C_2(\mathcal{T})-C_2(0)}{\mathcal{T}}=S(0),
\end{equation}
having used the property $c(t,t-\tau)=c(t-\tau,t)$, which follows from the definition.

\section{Creation of levitons}
\label{App:Levs_at_QPC}

A leviton and an anti-leviton can be created on a tight-binding chain by applying a time-dependent potential difference between two sides of the chain.\cite{Keeling2006} The time-dependent Hamiltonian reads
\begin{equation}
\hatH(t)=-\tb\sum_{m=1}^{M-1}(\ket{m}\bra{m+1}+\hc)+V(t)\sum_{m=1}^{L}\ket{m}\bra{m},
\label{eq:H_gate}
\end{equation}
where $\tb$ is the tunneling amplitude between neighboring sites and the potential drops between sites number $L<M$ and $L+1$ with $M$ being the number of sites. The time-dependent potential is chosen to be lorentzian,
\begin{equation}
V(t)=\frac{2\hbar\tau}{(t-t_e)^2+\tau^2},
\end{equation}
with width $\tau$, centered at $t=t_e$.

The Hamiltonian can be brought into an equivalent form by considering a continuum description of the potential,
\begin{equation}
\Phi(x,t)=\Theta (x_0-x)V(t).
\end{equation}
Here $x_0$ denotes the point along the $x$-axis, where the potential drops, and $\Theta(x)$ is the Heaviside step function. The potential can now be removed using a gauge transformation by choosing the gauge potential as
\begin{equation}
\gamma(x,t)=\int_{-\infty}^t dt'\Phi(x,t'),
\end{equation}
so that the transformed scalar potential $\Phi'(x,t)=\Phi(x,t)-\partial_t \gamma(x,t)$ is zero everywhere. This, however, leads to a non-zero transformed vector potential,
\begin{equation}
\mathbf{A}'(x,t)= [\partial_x \gamma(x,t)]\mathbf{x}=-\delta(x_0-x)\int_{-\infty}^t dt'V(t')\mathbf{x},
\nonumber
\end{equation}
where $\mathbf{x}$ is a unit vector along the $x$-axis. With this gauge transformation, the potential difference between the two sides of the chain has been removed. In turn, the presence of a vector potential should be included in the tight-binding Hamiltonian. This can be accomplished using a Peierls substitution\cite{Peierls1933} of the tunneling amplitudes
\begin{equation}
\tb\rightarrow \exp\of{-\frac{i}{\hbar}\int_{\ell}\mathbf{A}'(x,t)\cdot d\mathbf{x}}\tb,
\end{equation}
where the integral is evaluated along the line segment $\ell$ connecting the two neighboring sites. Inserting the vector potential, we see that only the tunneling amplitude where the potential drops should be modified as
\begin{equation}
\tb\rightarrow e^{i\phi(t)}\tb,
\end{equation}
with the time-dependent phase reading
\begin{equation}
\phi(t)=2\arctan\of{\frac{t-t_e}\tau}+\pi
\end{equation}
as in Eq.~(\ref{eq:phi_Lev}).

\end{document}